\documentclass{ws-procs9x6}

\begin{document}

\title{Deep Inelastic Scattering \\ from Light Nuclei}

\author{W. MELNITCHOUK$^1$, A. W. THOMAS$^2$}

\address{
$^1$	Jefferson Lab, 12000 Jefferson Avenue, Newport News, VA 23606 \\
$^2$	Department of Physics and Mathematical Physics, and Centre
	for the Subatomic Structure of Matter, University of Adelaide,
	5005, Australia}

%%%%%%%%%%%%%%%%%%%%%%%%%%%%%%%%%%%%%%%%%%%%%%%%%%%%%%%%%%%%%%
% You may repeat \author \address as often as necessary      %
%%%%%%%%%%%%%%%%%%%%%%%%%%%%%%%%%%%%%%%%%%%%%%%%%%%%%%%%%%%%%%

\maketitle

\abstracts{
We review recent developments in the study of deep inelastic scattering
from light nuclei, focusing in particular on deuterium, helium, and
lithium.
Understanding the nuclear effects in these systems is essential for the
extraction of information on the neutron structure function.
}

%%%%%%%%%%%%%%%%%%%%%%%%%%%%%%%%%%%%%%%%%%%%%%%%%%%%%%%%%%%%%%%%%%%%%%%%%
\section{Introduction}

For the past 20 years the nuclear EMC effect --- the nuclear medium
modification of the nucleon structure functions --- has posed a serious
challenge to models of the nucleus which involve quark degrees of freedom.
Although the basic features of the EMC effect can be understood within a
conventional nuclear physics framework, a quantitative description over
the entire range of Bjorken-$x$ appears beyond the capacity of a single
mechanism.
This has led to an assortment of nuclear models and effects which have
been postulated to account for the medium modifications of the structure
functions.

An impressive array of data on nuclear structure functions has by now
been accumulated from experiments at CERN, SLAC, DESY (Hermes), Fermilab,
Jefferson Lab and elsewhere, for a broad range of nuclei, and the
quality of recent data exploring extreme kinematics, both at low $x$ and
high $x$, is pushing nuclear models to their limits.
Surprisingly, the EMC effect in the lightest nuclei --- in particular the
$A=2$ and $A=3$ systems, where microscopic few-body calculations with
realistic potentials are more feasible --- is still unknown, leaving the
determination of the $A$ dependence of the effect incomplete.
In addition, the absence of data on the EMC effect in deuterium and helium
nuclei prevents the unambiguous determination of the structure function of
the free neutron, for which these nuclei are often used as effective
neutron targets.

Here we review the foundations of the conventional approach to deep
inelastic scattering (DIS) from few body nuclei, starting from a
covariant, relativistic framework.
We focus specifically on the intermediate and large $x$ region dominated
by valence quarks, and do not discuss the region of nuclear shadowing and
antishadowing at $x < 0.1$.
As examples, we describe the nuclear effects in spin-averaged and
spin-dependent structure functions of the deuteron, the $A=3$ mirror
nuclei, and lithium isotopes, paying particular attention to the nuclear
corrections which need to be applied when extracting neutron structure
functions.

%%%%%%%%%%%%%%%%%%%%%%%%%%%%%%%%%%%%%%%%%%%%%%%%%%%%%%%%%%%%%%%%%%%%%%%%%
\section{Formalism}

Away from the small-$x$ region ($x > 0.2 - 0.3$), DIS from a nucleus is
computed in the nuclear impulse approximation in terms of incoherent
scattering from individual bound nucleons in the nuclear
target.\cite{GST,MST}
The hadronic tensor for a nucleus $A$ can be written as\cite{MST,GL}
\begin{eqnarray}
W_{\mu\nu}^A(P,q)
&=& \int d^4p\ {\rm Tr}
\left[ \widehat {A}_{NA}(P,p) \cdot \widehat W^N_{\mu\nu}(p,q)
\right]\ ,
\end{eqnarray}
where $P$, $p$ and $q$ are the target nucleus, scattered nucleon, and
photon momenta, respectively.
For illustration we consider the case of spin-averaged DIS;
the generalization to spin-dependent scattering is
straightforward.\cite{DPOL_PLB,DPOL_PRC,DPOL_NR}
In this case, the most general expression for the truncated (off-shell)
nucleon tensor is given by\cite{MST}
\begin{eqnarray}
\widehat W_{\mu\nu}^N(p,q)
&=& g_{\mu\nu}
\left( I\ \widehat W_0\
    +\ \not\!p\ \widehat W_1\
    +\ \not\!q\ \widehat W_2
\right)\ ,
\end{eqnarray}
where the bound nucleon ``structure functions'' $\widehat W_{0,1,2}$
are functions of $p$ and $q$, and the (off-shell) nucleon-nucleus
amplitude can be written as
\begin{eqnarray}
\widehat {A}_{NA}(P,p) &=&
\left(  I\ {A}_{S}\
     +\ \gamma_\alpha\ {A}_V^{\alpha}
\right)\ ,
\end{eqnarray}
where a sum over residual $A-1$ nuclear states is implicit in the
functions $A_S$ and $A_V^{\alpha}$.
The spin-averaged $F_2$ structure function of the nucleus is
then given by
\begin{eqnarray}
F_2^A(x) &=&
\int d^4p\
\left(  {A}_{S} \widehat W_0\
     +\ p\cdot{A}_V \widehat W_1\
     +\ q\cdot{A}_V \widehat W_2
\right)\ .
\end{eqnarray}
In general the nuclear structure function cannot be written as a
one-dimensional convolution integral\cite{JAFFE} {\em even in the
impulse approximation}: factorization of amplitudes does not imply
factorization of structure functions.\cite{MST}
However, by taking either selective on-shell\cite{MST,MSTD} or
non-relativistic\cite{DPOL_NR,KPW} limits, one can identify a
convolution component, plus off-shell ($p^2\not= M^2$) corrections,
\begin{eqnarray}
\label{F2A}
F_2^A(x)
&=& \sum_N\ f_{N/A} \otimes F_2^N(x)\
 +\ \delta^{\rm (off)} F_2^A(x)\ ,
\end{eqnarray}
where $\otimes$ denotes convolution,
$f_{N/A} \otimes F_2^N (x) \equiv \int dy\ f_{N/A}(y)\ F_2^N(x/y)$,
with $y=p\cdot q/P\cdot q$ the light-cone fraction of the nucleus carried
by the interacting nucleon.

%%%%%%%%%%%%%%%%%%%%%%%%%%%%%%%%%%%%%%%%%%%%%%%%%%%%%%%%%%%%%%%%%%%%%%%%%
\section{Deuterium}

Being the simplest nuclear system, the deuteron has been studied most
extensively, both in non-relativistic and relativistic analyses.
Within the formalism described in Section 2, the deuteron $F_2$ structure
function can be calculated by smearing the nucleon structure functions
with the nucleon light-cone momentum distribution in the deuteron,
$f_{N/D}(y)$.
The off-shell corrections can be evaluated in terms of the relativistic
components of the deuteron wave functions and the $p^2$ dependence of the
truncated nucleon structure functions,\cite{MSTD} and are typically found
to be rather small,\
$|\delta^{\rm (off)} F_2^D| \stackrel{<}{\sim} 1\%$ for all $x$.

\begin{figure}[ht]
\centering{
\begin{picture}(30,150)(180,0)
\psfig{figure=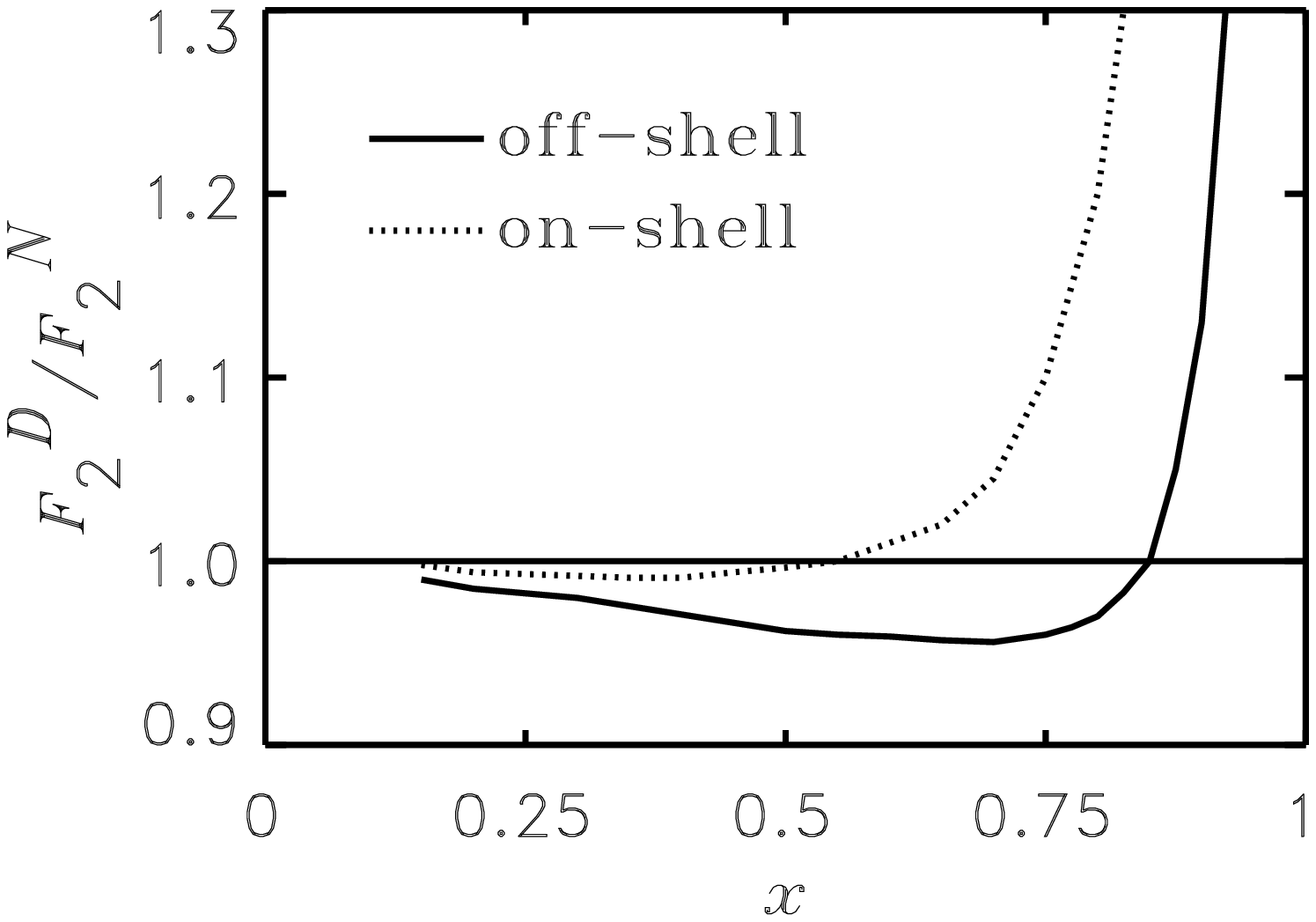,height=5.0cm}
\put(-20,0){\psfig{figure=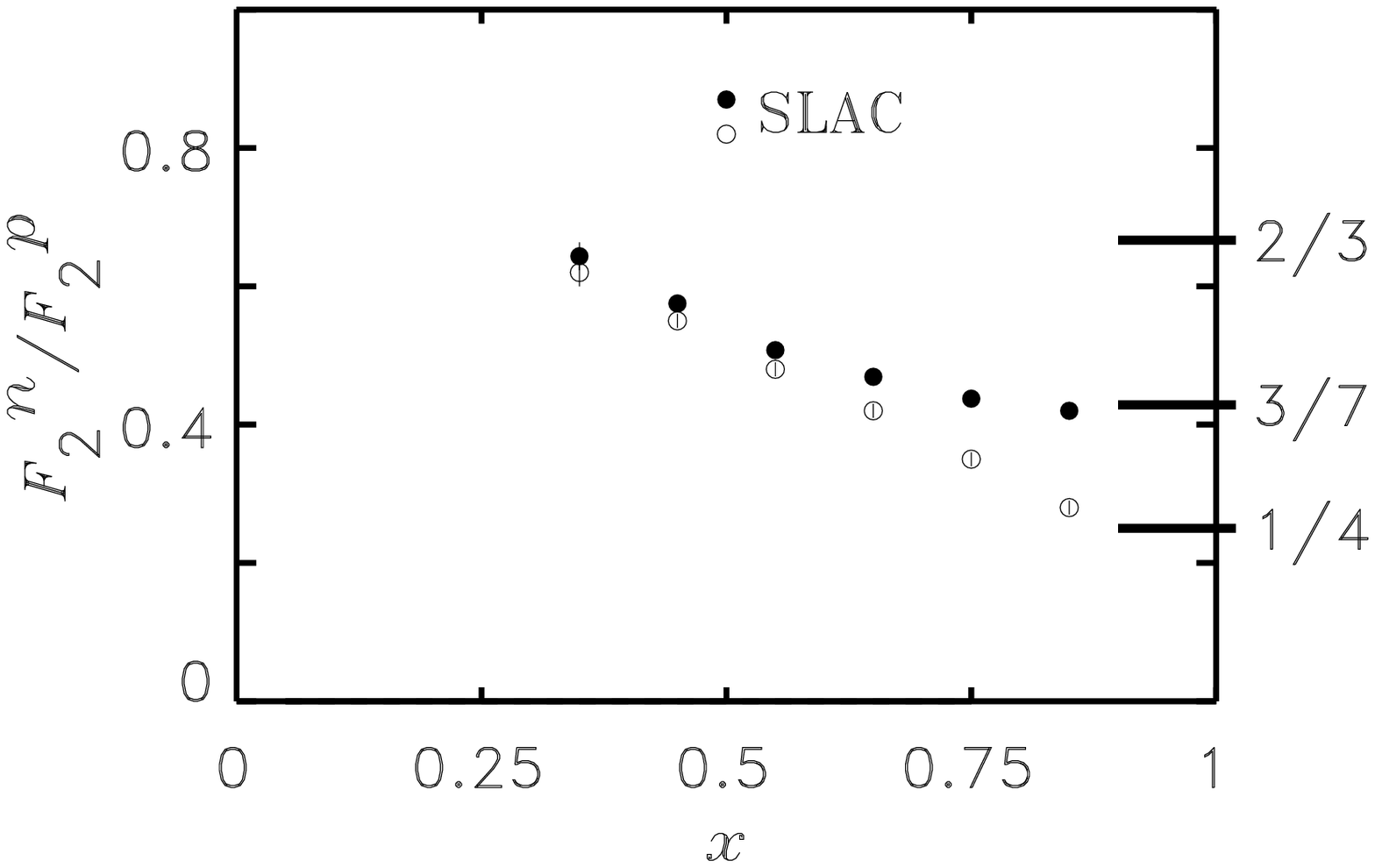,height=5.0cm}}
\end{picture}}
\caption{Left panel: ratio of the deuteron to nucleon $F_2$
	structure functions in the relativistic (off-shell)
	model\protect\cite{MSTD} (solid), and an on-shell
	model\protect\cite{FS} including Fermi motion only.
	Right panel: extracted neutron to proton ratio using
	the off-shell and on-shell models of the
	deuteron.\protect\cite{MT}}
\end{figure}

The calculated $F_2^D$ structure function for the relativistic model,
including the small off-shell correction, is shown in Fig.~1 (left panel)
as a ratio to the free nucleon structure function (solid curve).
Also shown is the result of an on-shell ansatz\cite{FS} in which
(effectively) only Fermi motion effects are included. This gives a much
smaller EMC effect in the deuteron.
{}From the measured deuteron and proton structure functions, and the
smearing function, $f_{N/D}(y)$, one can extract the structure function
of the neutron using the iterative deconvolution
procedure\cite{BODEK,KRAK} (which does not assume prior knowledge of
$F_2^n$).
Using $F_2^p$ and $F_2^D$ data from SLAC,\cite{WHITLOW} the resulting
$F_2^n/F_2^p$ ratio is shown in Fig.~1 (right panel) for the
relativistic\cite{MT,DEUT} and on-shell models (open and filled circles,
respectively).
The larger $F_2^n/F_2^p$ values in the former reflect a larger EMC
effect in the deuteron.
Despite the deuteron being a loosely bound system, the nuclear effects
at large $x$ can therefore play an important role in the extraction of
the free neutron structure function.

\begin{figure}[ht]
\centerline{\epsfxsize=8.0cm\epsfbox{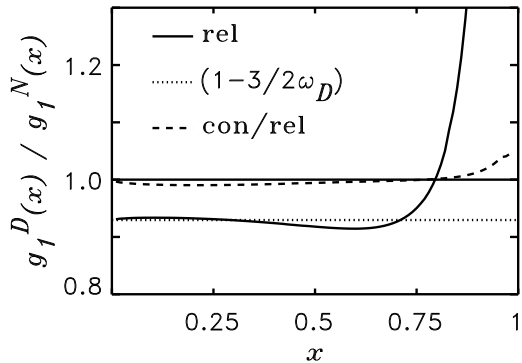}}
\caption{Ratio of deuteron and nucleon $g_1$ structure functions in the
	relativistic model\protect\cite{DPOL_PLB} (solid), and with a
	constant depolarization factor (dotted).
	Also shown is the ratio of the convolution component of
	$g_1^D$ to the total (dashed).}
\end{figure}

A similar analysis can be performed for the spin-dependent $g_1$
structure function of the deuteron, in terms of the spin-dependent
nucleon momentum distribution, $\Delta f_{N/D}(y)$, and relativistic
(off-shell) corrections,\cite{DPOL_PLB}
\begin{eqnarray}
g_1^D(x)
&=& \sum_N\ \Delta f_{N/D} \otimes g_1^N(x)\
 +\ \delta^{\rm (off)} g_1^D(x)\ ,
\end{eqnarray}
where $\Delta f_{N/D}$ is normalized to
$\int dy\ \Delta f_{N/D}(y) = 1-{3 \over 2} \omega_D$,
with $\omega_D$ the deuteron $D$ state probability.
The resulting ratio of the deuteron to nucleon $g_1$ structure functions
is illustrated in Fig.~2.
Also shown is the result assuming
$\delta^{\rm (off)} g_1^D \rightarrow 0$ and
$\Delta f_{N/D}(y) \rightarrow (1-{3 \over 2} \omega_D) \delta(1-y)$,
in which the ratio is given by a constant depolarization factor for all $x$.
For $x < 0.7$ the nuclear effects are weakly dependent on $x$, so that
the use of a constant depolarization factor is a reasonable approximation,
although it breaks down dramatically for $x > 0.75$.
Furthermore, the convolution approximation is quite accurate for
$x < 0.8$, with $\stackrel{<}{\sim} 1-2\%$ deviations from the full
result.
The largest uncertainty in the extraction of $g_1^n$ over this range of
$x$ is due to uncertainty in the $D$-state probability.

%%%%%%%%%%%%%%%%%%%%%%%%%%%%%%%%%%%%%%%%%%%%%%%%%%%%%%%%%%%%%%%%%%%%%%%%%
\section{$A=3$ Nuclei}

As the lightest mirror nuclei, $^3$He and $^3$H provide a unique
laboratory for studying nuclear effects in structure functions, since
many of their properties are similar up to nuclear charge symmetry
breaking effects.
Furthermore, since the doubly represented nucleons tend to couple to
spin 0, the spin of the $A=3$ nuclei is carried largely by the singly
represented nucleon ($n$ in the case of $^3$He, $p$ in the case of $^3$H).

At present the size of the nuclear EMC effect in $^3$H is not known, and
only scant data exist\cite{HERMES} on the $F_2$ structure function of
$^3$He at lower $Q^2$.
The predicted nuclear EMC ratio for $^3$He,
\begin{eqnarray}
R(^3{\rm He}) &=& { F_2^{^3{\rm He}} \over 2 F_2^p + F_2^n}\ ,
\end{eqnarray}
is plotted in Fig.~3 (left panel), where $F_2^{^3{\rm He}}$ is evaluated
as in Eq.~(\ref{F2A}) via a convolution of the bound $p$ and $n$
structure functions with the nucleon momentum distributions in $^3$He,
$f_{N/^3{\rm He}}(y)$.
The latter are obtained from three-body wave functions calculated by
solving the Faddeev equations with a modified Paris potential\cite{BAT}
(dashed) and via a variational approach\cite{ROME,CL} (dot-dashed).
Since relativistic wave functions of $A=3$ nuclei are not yet available,
we cannot quantify the off-shell correction,
$\delta^{({\rm off})} F_2^{^3{\rm He}}$.
However, analysis of quasi-elastic scattering from $^4$He, in which
binding effects are larger than those in $^3$He, suggests\cite{MTT}
rather small modifications in the intrinsic nucleon structure
($< 5\%$ for $x < 0.8$).
The theoretical ratios are compared with data from HERMES\cite{HERMES}
on the ratio of $^3$He to $D+p$ structure functions, and agree quite
well within the current errors.\cite{LONG}
The effect on $R(^3{\rm He})$ of including the EMC effect in the
deuteron (solid) is relatively small for $x < 0.8$.

\begin{figure}[ht]
\centering{
\begin{picture}(10,150)(180,0)
\psfig{figure=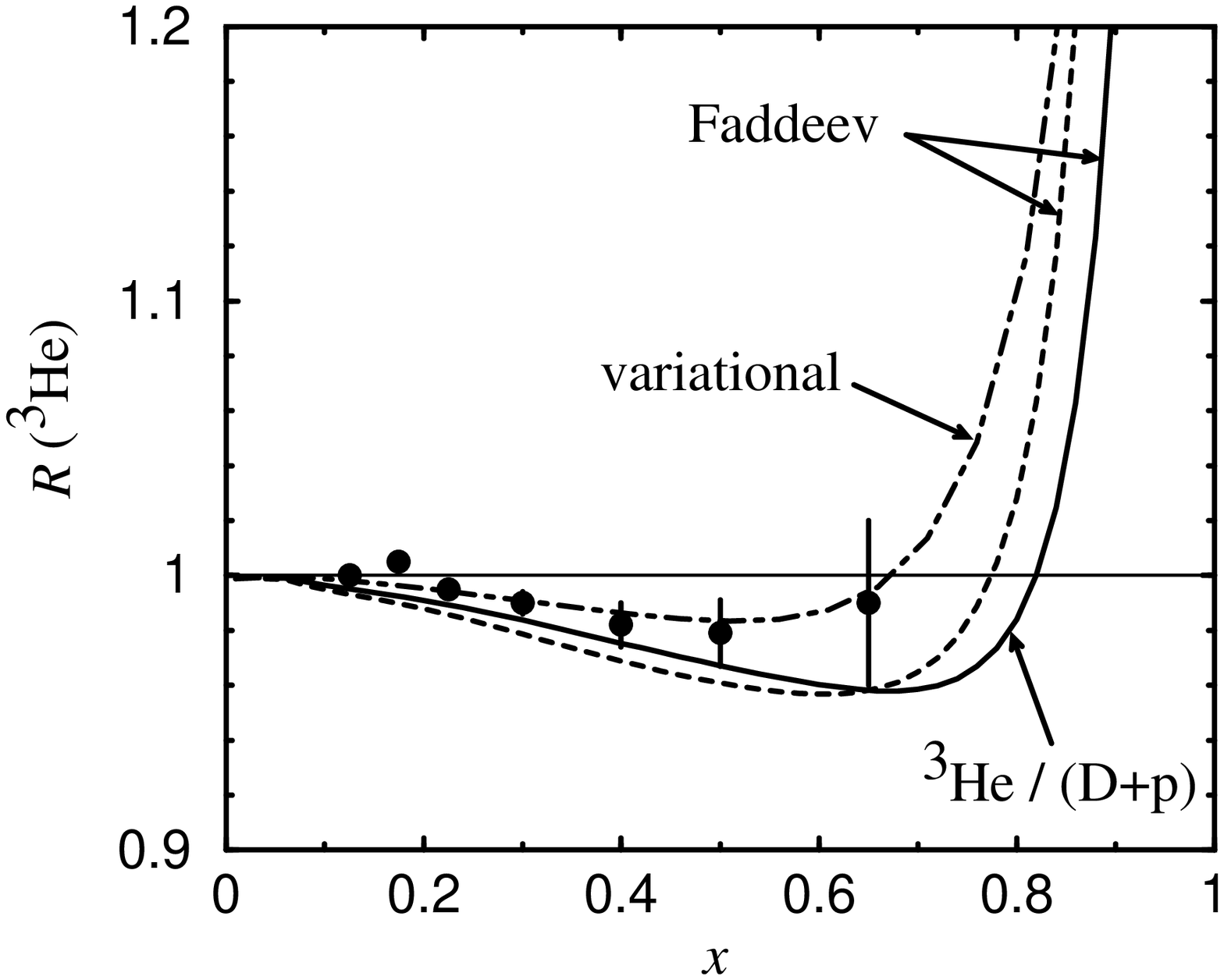,height=4.8cm}
\put(10,-4){\psfig{figure=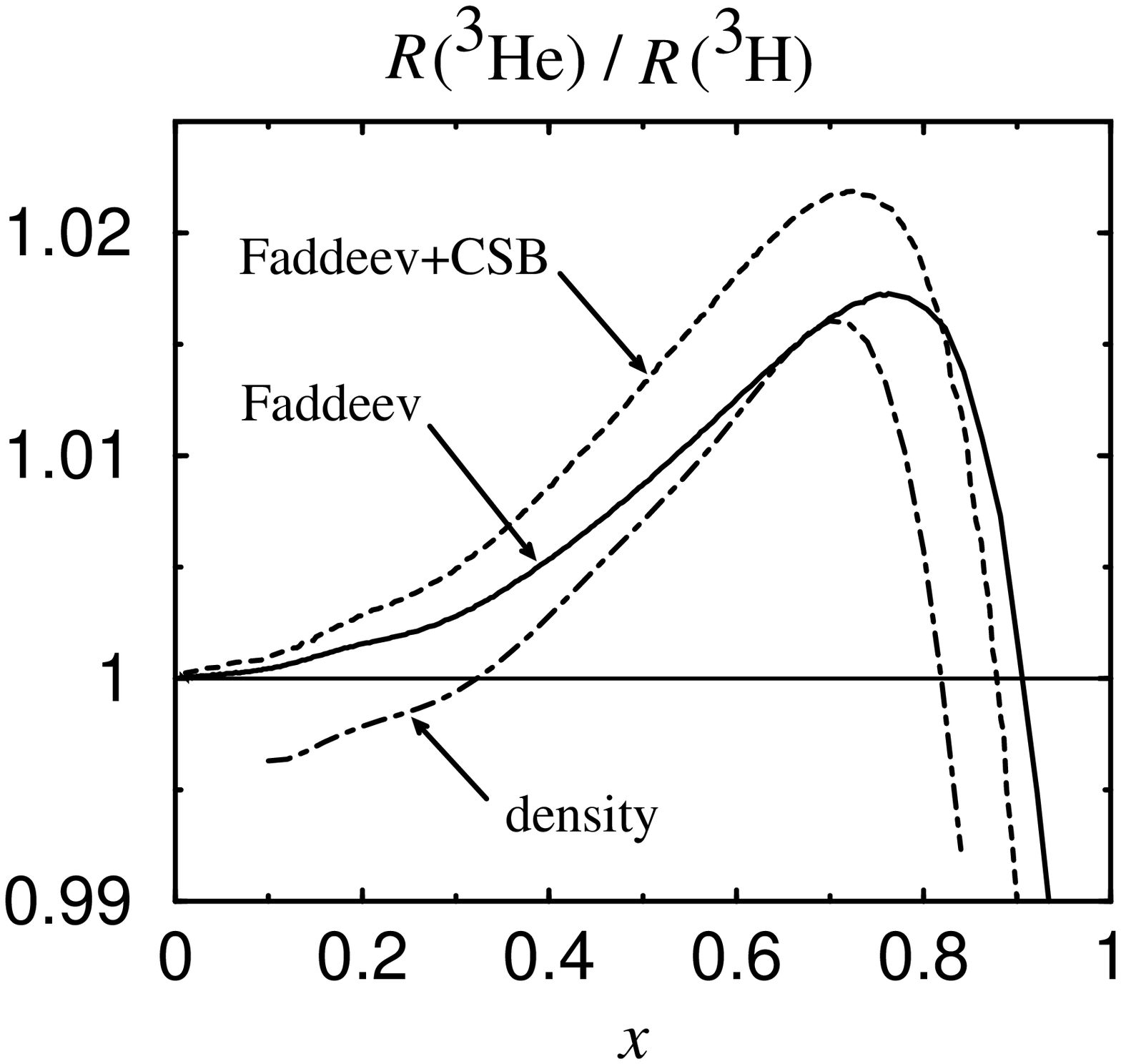,height=5.5cm}}
\end{picture}}
\caption{Left panel: nuclear EMC ratio in $^3$He, compared with data on
	the $^3$He/$(D+p)$ ratio.\protect\cite{HERMES}
	The effect of nuclear corrections in the deuteron is indicated
	by the solid curve.
	Right panel: ratio of EMC ratios in $^3$He and $^3$H,
	for the Faddeev wave function model (solid), including explicit
	charge symmetry breaking effects (dashed),\protect\cite{AFN}
	and an ansatz based on the nuclear density extrapolation
	model.\protect\cite{COMMENT}}
\end{figure}

While the model dependence of the EMC effect in $^3$He is at the few \%
level, the fact that $^3$He and $^3$H are mirror nuclei means that one
can expect similar medium modification in both nuclei.
This is illustrated in Fig.~3 (right panel), where the ratio of EMC
ratios for $^3$He and $^3$H,
${\bf R} \equiv R(^3{\rm He})/R(^3{\rm H})$,
is plotted for the standard Faddeev wave function (solid),\cite{AFN}
and a modified wave function including effects of charge symmetry
breaking (dashed) which accounts for the difference in binding energies
of $^3$He and $^3$H.
For comparison, the result of the density extrapolation model is also
shown.
Even though this model\cite{FS} predicts a difference in the absolute
size of the EMC effect between $^3$He and $^3$H of order 40\%, it
nevertheless gives rise to a similar effect in the ratio
${\bf R}$.\cite{COMMENT}
{}From the calculated ratio ${\bf R}$ and a measurement of the ratio of
the $^3$He and $^3$H structure functions, one can directly extract the
neutron to proton ratio,\cite{AFN}
\begin{eqnarray}
{ F_2^n \over F_2^p }
&=& { 2 {\bf R} - F_2^{^3{\rm He}}/F_2^{^3{\rm H}}
\over 2 F_2^{^3{\rm He}}/F_2^{^3{\rm H}} - {\bf R} }\ ,
\end{eqnarray}
with significantly less sensitivity to nuclear effects than in Fig.~1
using the inclusive deuteron data.\cite{AFN,OTHER}

Turning to the spin polarized $A=3$ system, the fact that in $^3$He the
$pp$ pair couples predominantly to spin 0 allows polarized $^3$He to be
used as an effective polarized neutron target\cite{BLANK,FRIAR}.
The $g_1$ structure of $^3$He can be written as
\begin{eqnarray} 
g_1^{^3{\rm He}}(x)
&\approx& 2 \Delta f_{p/^3{\rm He}} \otimes g_1^p(x)\
       +\   \Delta f_{n/^3{\rm He}} \otimes g_1^n(x)\ ,
\end{eqnarray} 
where the spin-dependent nucleon momentum distributions in $^3$He are
normalized such that\cite{BAT,CIOFI}
$\int dy\ \Delta f_{n/^3{\rm He}}(y) \equiv \rho_n \approx 88\%$ and
$\int dy\ \Delta f_{p/^3{\rm He}}(y) \equiv \rho_p \approx -2\%$.
The similarity between the $^3$He and neutron $g_1$ structure functions
is illustrated in Fig.~4 (left panel), reflecting the small proton
contribution to the $^3$He polarization.
For a polarized $^3$H target, the difference between the $g_1$ structure
function of $^3$H,
\begin{eqnarray}
g_1^{^3{\rm H}}(x)
&\approx& \Delta f_{p/^3{\rm H}} \otimes g_1^p(x)\ ,
\end{eqnarray}
and $g_1^p$ would be even smaller because, in addition to the small size
of the neutron polarization in $\vec{^3{\rm H}}$,
the intrinsic magnitude of $g_1^n$ is much smaller than that of $g_1^p$.
A measurement of $g_1^{^3{\rm H}}$ would therefore allow an accurate
determination of the
$\Delta f_{p/^3{\rm H}} \approx \Delta f_{n/^3{\rm He}}$ distribution,
which could then be used to minimize the uncertainty in the extraction of
$g_1^n$ from $g_1^{^3{\rm He}}$.
The effects of possible $\Delta$ resonance components in the $^3$He wave
function have also been examined by Bissey {\em et al.}\cite{BGST}

\begin{figure}[ht]
\centering{
\begin{picture}(10,150)(180,0)
\psfig{figure=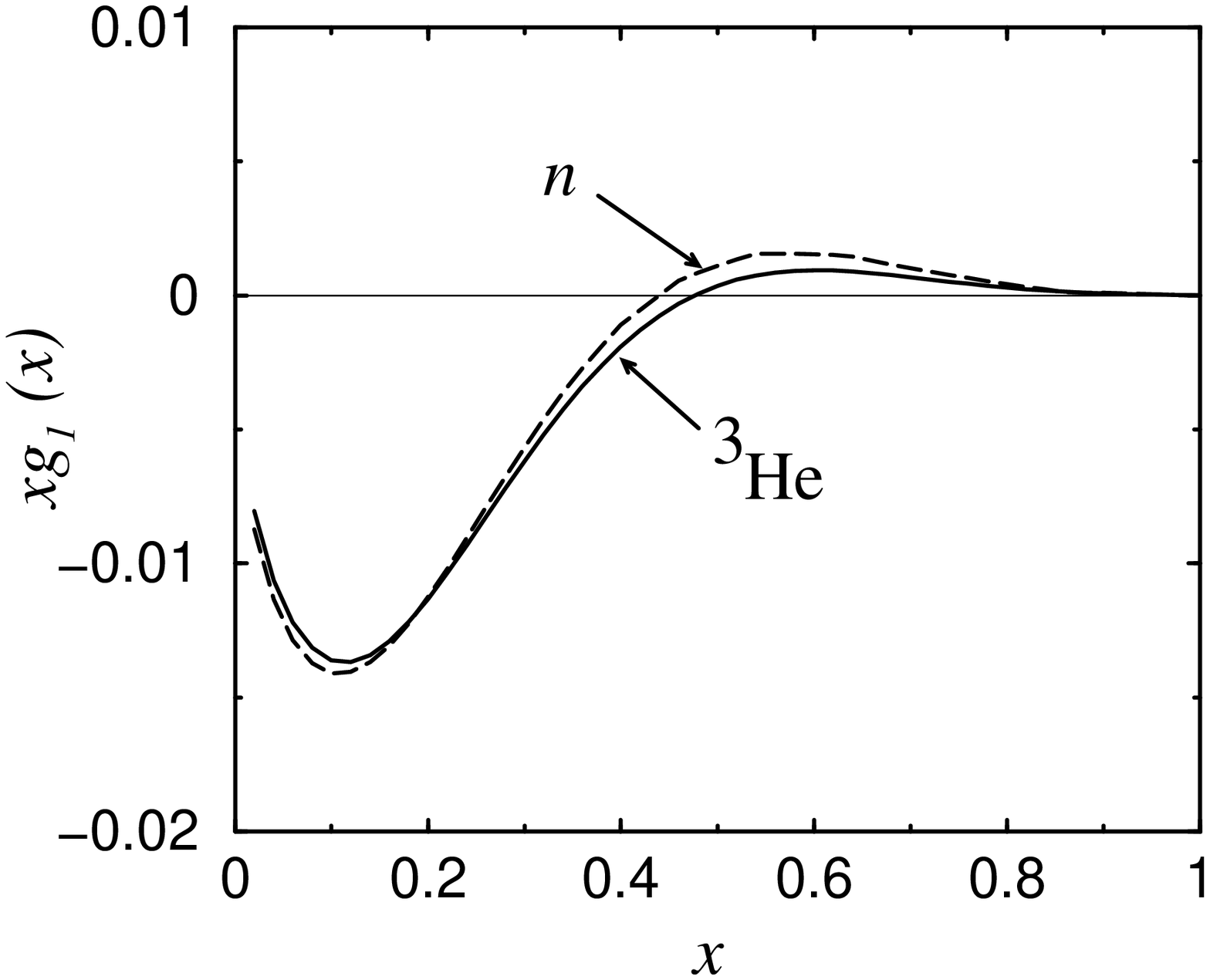,height=5.0cm}
\put(0,0){\psfig{figure=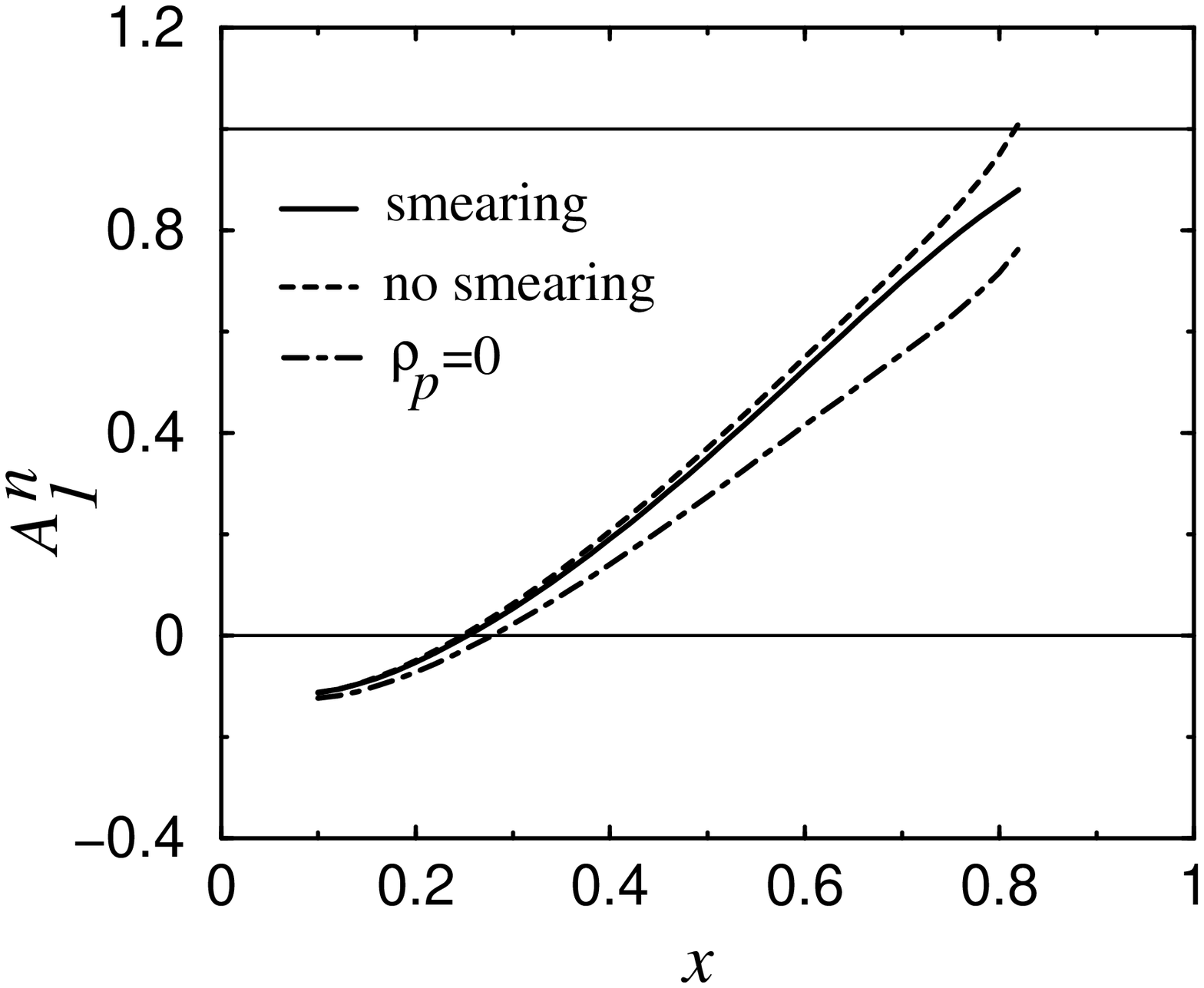,height=5.0cm}}
\end{picture}}
\caption{Left panel: $x g_1$ structure function of the neutron and $^3$He.
	Right panel: neutron polarization asymmetry $A_1^n$ obtained from
	the $^3$He asymmetry including smearing corrections (solid),
	without smearing (dashed), and neglecting the proton polarization,
	$\rho_p=0$ (dot-dashed).}
\end{figure}

Because the $g_1^n$ structure function changes sign at intermediate $x$,
it is not meaningful to calculate an EMC-type ratio.
On the other hand, taking the ratio of the $g_1$ to $F_1$ structure
functions (where $F_1 \approx F_2/2x$ at large $Q^2$), one can better
visualize the nuclear corrections in the polarization asymmetry,
\begin{eqnarray}
A_1(x) &\approx& { g_1(x) \over F_1(x) }\ ,
\end{eqnarray}
which is the quantity measured directly in polarized DIS experiments.
In Fig.~4 (right panel) the $A_1^n$ asymmetry is shown, extracted from
the $^3$He asymmetry including smearing corrections (solid).
For comparison, the result without smearing is also shown (dashed)
--- in this case the distributions
$\Delta f_{N/^3{\rm He}}(y) \rightarrow \rho_N \delta(1-y)$, and the
shapes of $g_1^{^3{\rm He}}$ and $g_1^n$ are assumed to be the same.
This appears to be a reasonable approximation for $x < 0.7$, but breaks
down beyond $x \sim 0.8$.
Over the same region of $x$ the proton polarization is also seen to play
a relatively minor role, although it must be included if precision at the
level of a few percent is required.

An accurate determination of nuclear effects is also necessary for the
extraction of the $g_2$ structure function of the neutron.
The $g_2$ structure function provides the cleanest means to access
information on higher twist contributions in polarized DIS.
Of particular importance is the $x^2$-weighted moment of $g_2$, which
enters in the twist-3 $d_2$ matrix element.
The nuclear corrections to the neutron $g_2$ structure function are
illustrated in Fig.~5, where the $x^2$-weighting emphasizes the
larger-$x$ region.
Since there is practically no information on the twist-3 component of
$g_2$, the calculation has been performed assuming twist-2 dominance of
$g_2^n$.
The effect of smearing appears to be small, although not negligible,
and needs to be included for an accurate determination of $g_2^n$ at
large $x$.

\begin{figure}[ht]
\centerline{\epsfxsize=6.5cm\epsfbox{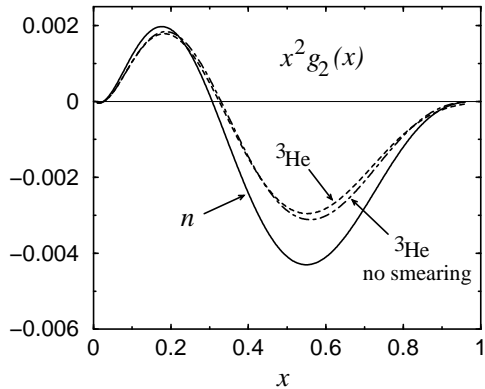}}
\caption{$x^2$-weighted $g_2$ structure function of the neutron (solid),
	and of $^3$He, including smearing corrections (dashed),
	and without smearing (dot-dashed).}
\end{figure}

%%%%%%%%%%%%%%%%%%%%%%%%%%%%%%%%%%%%%%%%%%%%%%%%%%%%%%%%%%%%%%%%%%%%%%%%%
\section{Lithium}

$\beta$-unstable Li isotopes provide a promising new avenue for exploring
the quark structure of few body nuclei.
The $^{11}$Li nucleus, which is located at the neutron drip line, is
believed to be well approximated as a $^9$Li core plus two weakly bound
valence neutrons, which form a halo structure.
Because of the weak binding, it has been suggested\cite{SUTT} that the
neutron momentum distribution in $^{11}$Li should be very sharp and
symmetric around $y = 1$, making it an excellent candidate for an
effective neutron target.
Polarized $^6$Li nuclei, which have with spin 1 and isospin 0, have been
suggested as effective polarized ($\approx 86\%$) deuteron targets, in
the approximation that $\vec{^6{\rm Li}} \approx \alpha + \vec{D}$.
Furthermore, polarized $^{7,9,11}$Li isotopes can also be used to study
the novel structure function\cite{SUTT,JM} associated with spin 3/2
systems, $^{3/2}_3g_1$.
Deep inelastic scattering from the $^7$Li--$^7$Be mirror nuclei can also
be used to explore the medium modifications of the Bjorken and Gottfried
sum rules.\cite{SUTT,GS}
Studies with these nuclei may one day be undertaken at radioactive beam
facilities, such as those proposed at RIKEN or GSI.
The merits of using other nuclear targets, such as nitrogen, have been
elaborated by Rondon.\cite{RONDON}

%%%%%%%%%%%%%%%%%%%%%%%%%%%%%%%%%%%%%%%%%%%%%%%%%%%%%%%%%%%%%%%%%%%%%%%%%
\section{Conclusions}

The unprecedented quality of recent data on few body nuclear structure
functions from new generations of accelerators is demanding better
understanding of nuclei in extreme kinematic regions, such as at large
$x$, where the effects of relativity and nucleon substructure play a
more prominent role.
This challenge is being met by concurrent progress being made in the
theory of DIS from few body nuclei, especially deuterium and $A=3$
nuclei.

In addition to more accurately determining the response of these nuclei
to electromagnetic probes, an important practical necessity lies in
controlling the nuclear corrections when extracting information on the
structure of the neutron, for which light nuclei are often used as
effective neutron targets.
While much of the focus has been on the $F_2$ structure function, there
is a growing need to understand the nuclear effects in other structure
functions, such as $g_1$ and $g_2$, as well as the new structure
functions available for higher spin nuclei --- both in the deep inelastic
and resonance regions.
The interplay between theoretical developments and anticipated future
data promises to provide even deeper insights into the quark structure
of nuclei.

%%%%%%%%%%%%%%%%%%%%%%%%%%%%%%%%%%%%%%%%%%%%%%%%%%%%%%%%%%%%%%%%%%%%%%%%%
\section*{Acknowledgments}

We are grateful to F.~Bissey and S.~Liuti for providing their light-cone
momentum distribution functions.
This work was supported by the U.S. Department of Energy contract
\mbox{DE-AC05-84ER40150}, under which the Southeastern Universities
Research Association (SURA) operates the Thomas Jefferson National
Accelerator Facility (Jefferson Lab) and by the Australian Research
Council.

\newpage
%%%%%%%%%%%%%%%%%%%%%%%%%%%%%%%%%%%%%%%%%%%%%%%%%%%%%%%%%%%%%%%%%%%%%%%%%

\end{document}